\documentclass[12pt]{elsarticle}

\pdfoutput=1

\usepackage{amsmath,bm,amsfonts,amssymb,mathrsfs,mathtools} 
\usepackage{lineno}
\usepackage[utf8]{inputenc} 
\usepackage[T1]{fontenc}    
\usepackage{hyperref}       
\usepackage{url}            
\usepackage{booktabs}       
\usepackage{amsfonts}       
\usepackage{nicefrac}       
\usepackage{microtype}      
\usepackage{lipsum}		
\usepackage{graphicx}
\usepackage{natbib}
\usepackage{doi}
\usepackage{subcaption}
\graphicspath{{figures}}

\newcommand{\bv}[1]{\mathbf{#1}}
\newcommand{\bg}[1]{\boldsymbol{#1}}
\newcommand{\giv}[1][]{\:#1\vert\:}
\newcommand{\ex}{\mathbb{E}}
\newcommand{\var}{\mathbb{V}}
\newcommand{\data}{\mathcal{D}}

\renewcommand{\d}[1]{\ensuremath{\operatorname{d}\!{#1}}}
\DeclarePairedDelimiterX{\infdivx}[2]{(}{)}{%
	#1\;\delimsize\|\;#2%
}
\newcommand{\kldiv}{D_{\text{KL}}\infdivx}
\newcommand{\jsdiv}{D_{\text{JSD}}\infdivx}
\newcommand{\Sgmm}{{\boldsymbol{\mathcal{S}}}}

\DeclareMathOperator*{\argmin}{arg\,min}
\DeclareMathOperator{\arcsinh}{arcsinh}

\usepackage{xcolor}
\usepackage[normalem]{ulem} 

\newcommand{\mub}{{\boldsymbol{\mu}}}

\newcommand{\pib}{{\boldsymbol{\pi}}}
\newcommand{\thetab}{{\boldsymbol{\theta}}}

\newcommand{\epsilonb}{\boldsymbol{\epsilon}}

\newcommand{\Sigmab}{\boldsymbol{\Sigma}}

\newcommand{\xb}{\mathbf{x}}

\newcommand{\vb}{\mathbf{v}}

\newcommand{\ub}{\mathbf{u}}
\newcommand{\zb}{\mathbf{z}}
\newcommand{\Ab}{\mathbf{A}}

\newcommand{\Ub}{\mathbf{U}}

\newcommand{\Hb}{\mathbf{H}}

\newcommand{\Xb}{\mathbf{X}}

\date{}

\journal{}

\begin{document}

\begin{frontmatter}

\title{Robust scalable initialization for Bayesian variational inference with multi-modal Laplace approximations}

\address[label1]{Sandia National Laboratories, Livermore, California 94550, USA}

\author[label1]{Wyatt Bridgman}

\author[label1]{Reese Jones}

\author[label1]{Mohammad Khalil}

\begin{abstract}
For predictive modeling relying on Bayesian model calibration, fully independent, or ``mean-field'', Gaussian distributions are often used as approximate probability density functions in variational inference since the number of variational parameters grows only linearly with the number of unknown model parameters. The resulting diagonal covariance structure coupled with unimodal behavior can be too restrictive to provide useful approximations of intractable Bayesian posteriors exhibiting highly non-Gaussian behavior, including multimodality. High-fidelity surrogate posteriors for these problems can be obtained by considering the family of Gaussian mixtures.
Gaussian mixtures are capable of capturing multiple modes and approximating any distribution to an arbitrary degree of accuracy while maintaining some analytical tractability. 
Variational inference with Gaussian mixtures with full-covariance structures suffers from a quadratic growth in variational parameters with the number of model parameters.
Coupled with the existence of multiple local minima due to strong nonconvex trends in the loss functions often associated with variational inference, these challenges motivate the need for robust initialization procedures to improve the performance and computational scalability of variational inference with mixture models.

In this work, we propose a method for constructing an initial Gaussian mixture model approximation that can be used to warm-start the iterative solvers for variational inference.
The procedure begins with a global optimization stage in model parameter space in which local gradient-based optimization, globalized through multistart, is used to determine a set of local maxima, which we take to approximate the mixture component centers.
Around each mode, a local Gaussian approximation is constructed via the Laplace approximation.
Finally, the mixture weights are determined through constrained least squares regression.
The robustness and scalability of the proposed methodology is demonstrated through application to an ensemble of synthetic tests using high-dimensional, multimodal probability density functions. Finally, the approach is demonstrated with an inversion problem in structural dynamics involving unknown viscous damping coefficients.
\end{abstract}


\end{frontmatter}

\section{Introduction}\label{sec:introduction}
A frequent problem arising in statistical model calibration is the approximation of intractable density kernels resulting from Bayesian inference. For these problems, a popular approach is to use a method such as Markov chain Monte Carlo (MCMC) \cite{brooksMarkovChainMonte1998,andrieuParticleMarkovChain2010} that provide samples distributed according to the target posterior PDF using a carefully constructed Markov Chain. This approach suffers from scalability issues due to being inherently sequential and can display slow convergence rates for high-dimensional distributions \cite{van2018simple}. Dropout  \cite{galDropoutBayesianApproximation2016} provides a more scalable sampling strategy for posteriors in the context of large neural networks and proceeds by repeated stochastic modulations of the weights in the network and evaluating the resulting perturbed model. Dropout also has a theoretical foundation as a Variational Inference approximation to a Deep Gaussian process  \cite{damianouDeepGaussianProcesses2013}.

An alternative strategy to sampling techniques is Variational Inference (VI) \cite{bleiVariationalInferenceReview2017} which approximates an intractable posterior PDF using a parametric family of densities. VI recasts approximate inference as an optimization problem, which allows for iterative techniques such as gradient descent to be applied \cite{blundellWeightUncertaintyNeural2015}. It can offer better scalability than some sampling approaches, such as MCMC, for certain parametric densities. A common choice is Mean Field Variational Inference (MFVI) in which employs a multivariate Gaussian with a diagonal covariance to limit the number of variational parameters to only twice the number of unknown model parameters. Both MFVI and Dropout have limited expressiveness \cite{foongExpressivenessApproximateInference2020}. MFVI tends to underestimate the uncertainty of the posterior \cite{Han2019StatisticalII}, while Dropout has been shown to perform similarly to MFVI for uncertainty quantification in machine learning problems \cite{ngEstimatingUncertaintyNeural2022}. 

Approximation of non-Gaussian, multimodal posterior PDFs is an important research task as these can arise in the context of nonlinear, many-parameter models and with sparse and/or noisy data across many different fields of application \cite{rafteryEstimatingProjectingTrends2010,jonoskaIncrementalMixtureImportance2017,ferozMultiNestEfficientRobust2008,rodriguezMultimodalWaterAge2020,zhangHighprecisionProbabilisticUncertainty2019}. Probability densities displaying multimodal  behavior present a particular challenging case for each of the aforementioned methods. Sampling strategies exhibit difficulties sampling across multiple modes \cite{yaoStackingNonmixingBayesian2020,ferozMultiNestEfficientRobust2008} while MFVI is limited to a unimodal approximation. To obtain better approximations to the posterior, higher-fidelity distributions such as full-covariance Gaussians or Gaussian Mixture Models (GMMs) can be used but suffer from poor scalability due to the quadratic growth in the number of variational parameters.  Objective functions for VI, such as the evidence lower bound (ELBO), also often display strong non-convex trends leading to optimizer getting stuck in poor local minima \cite{kingmaIntroductionVariationalAutoencoders2019}, an issue that can be alleviated through globalization strategies \cite{bowmanGeneratingSentencesContinuous2016,sonderbyHowTrainDeep2016} and effective initialization \cite{rossiGoodInitializationsVariational2019}.

In this work, we develop a global optimization and Laplace approximation (GOLA)  procedure that addresses the foregoing difficulties in obtaining high-fidelity approximations to posteriors by forming an ensemble of local models. Such ensembles form Gaussian approximations at multiple modes of the posterior where the weights of the components are determined through constrained linear regression. This method provides a GMM approximation at low cost compared to VI. GOLA is shown to be an effective initialization strategy for VI with GMMs as well as a possible alternative approximation when VI is too expensive to carry out. The proposed strategy leverages the growing body of literature investigating the theoretical foundation of Laplace approximations (LA) \cite{immerImprovingPredictionsBayesian2021,khanApproximateInferenceTurns2020} and showing that the LA performs well in a variety of machine learning with uncertainty quantification (UQ) applications  \cite{ritterScalableLaplaceApproximation2018,immerScalableMarginalLikelihood2021,ritterOnlineStructuredLaplace2018,daxbergerBayesianDeepLearning2022,daxbergerLaplaceReduxEffortless2021}. 

Repeated LAs have also been used to construct GMM approximations to intractable posteriors in Ref.\cite{bornkampApproximatingProbabilityDensities2011} by iterating on the residual between the current GMM approximation and posterior. Gaussian components are added around discovered modes of the residual using the LA. While this approach can theoretically achieve arbitrarily small approximation errors, it is inherently sequential and each iteration increases the computational complexity by adding terms to the residual.  Alternative methods for constructing GMM approximations include using iterative VI -based algorithms \cite{guoBoostingVariationalInference2016, millerVariationalBoostingIteratively2016} and importance sampling approaches where components are continually added based on some convergence criterion \cite{kurtzCrossentropybasedAdaptiveImportance2013,cappeAdaptiveImportanceSampling2008,hoogerheideClassAdaptiveImportance2012,khorunzhinaFiniteGaussianMixture2019,hesterbergWeightedAverageImportance1995,steeleComputingNormalizingConstants2006,rafteryEstimatingProjectingTrends2010} including a procedure that involves a global optimization stage \cite{jonoskaIncrementalMixtureImportance2017}. Other strategies involve clustering \cite{giordaniAdaptiveIndependentMetropolis2010} and using normalizing flows to obtain a mixture model with richer covariance structure \cite{liuVariationalInferenceGaussian2019}. The majority of methods discussed above will suffer from scalability issues in high-dimensional settings because of steps that involve optimization over the full set of mixture parameters or integral approximations via quadrature. In addition, many of the techniques for discovering new modes employed by these methods tend to search only in a local vicinity of modes already discovered. The proposed GOLA approach can carry out global optimization in parallel and achieve better scalability at the cost of missing non-Gaussian behavior around the modes.

The remainder of the paper is organized as follows: section \ref{sec:methods} describes the methods used, section \ref{sec:results} contains results with \ref{sec:robustness} and \ref{sec:scalability} describing an analysis of the robustness and scalability of the GOLA method and \ref{sec:physics-application} providing a physics-based, structural dynamics exemplar to demonstrate how the method performs in practice.

\section{Methods}\label{sec:methods}
In this section, variational inference is described along with optimization techniques used to carry it out in practice. Following this, a detailed description of the proposed GOLA method is provided and summarized in algorithm form.

\subsection{Variational Inference}
Given an intractable density resulting from Bayesian inference
\begin{equation}
	p(\bv{z} \giv \data) = \frac{p(\data \giv \bv{z})p(\bv{z})}{p(\data)} = \frac{p(\data \giv \bv{z})p(\bv{z})}{\int p(\data \giv \bv{z}) p(\bv{z}) \d{\bv{z}}}
\end{equation}
VI seeks an approximating distribution $q_{\bm{\theta}}(\bv{z}) \in \mathcal{F}_{\bm{\theta}}$ in some parametric family $\mathcal{F}_{\bm{\theta}}$ by minimizing an error measure such as the Kullback-Liebler (KL) divergence
\begin{equation}
	q_{\bm{\theta}}(\bv{z}) = \argmin_\thetab \kldiv{q_{\bm{\theta}}(\bv{z})}{p(\bv{z} \giv \data)}
\end{equation}

The optimization problem of minimizing the KL-divergence is often reformulated as an equivalent optimization problem of minimizing the negative Evidence Lower Bound (ELBO) \cite{kingmaIntroductionVariationalAutoencoders2019}
\begin{equation}
	q_{\bm{\theta}}(\bv{z}) = \argmin_{\thetab} \kldiv{q_{\bm{\theta}} (\bv{z})}{p(\bv{z})} - \ex_{q_{\bm{\theta}} (\bv{z})}  \left[ \log p(\data \giv \bv{z}) \right] 
\end{equation}
where the optimization often proceeds using gradient-based schemes. By expressing the negative ELBO as $\ex_{q_{\bg{\theta}} (\bv{z})} \left[ \log \frac{q_\thetab(\zb)}{p(\zb)} - \log p(\data \giv \zb) \right] = \ex_{q_{\bg{\theta}} (\bv{z})} \left[ f(\zb) \right]$, a gradient estimator to drive the minimization can be formed by representing the latent random variables as a transformation $\bv{z} = t_{\bg{\theta}}(\bg{\epsilon})$ of another random variable $ \bg{\epsilon} \sim p(\bg{\epsilon})$ that depends deterministically on the parameters $\bg{\theta}$ such that
\begin{equation} \label{eq:reparametrization}
	\nabla_\thetab \ex_{q_\thetab(\zb)} \left[ f(\zb) \right] = \ex_{p(\epsilonb)} \left[ \nabla_\thetab f(\zb) \right]
\end{equation}
which can be estimated using straightforward Monte Carlo techniques. Obtaining such reparametrization gradients are more difficult but possible for mixture models \cite{figurnovImplicitReparameterizationGradients2019,gravesStochasticBackpropagationMixture2016}.

The score function provides another unbiased estimator of the ELBO gradient in the form
\begin{equation}
		\nabla_\thetab \ex_{q_\thetab(\zb)} \left[ f(\zb) \right] = \ex_{q_\thetab(\zb)} \left[ f(\zb)  \nabla_\thetab \log q_\thetab(\zb) \right] 
\end{equation}
where the derivative acts on the surrogate posterior PDF with respect to its parameters. 
As this derivative is often easily computed, this estimator has broader applicability than the reparametrization, Eq. \ref{eq:reparametrization}, at the expense of higher variance.

\subsection{Global optimization and Laplace approximation method}

To capture the non-Gaussian trends in posterior PDFs normally encountered in nonlinear-in-parameter models, such as Neural Networks (NNs), we propose a method that seeks an approximation $q_{\bg{\theta}}(\bv{z})$ to $p(\bv{z} \giv \data)$ in the form of a Gaussian mixture model
\begin{equation}
	q_{\bg{\theta}}(\bv{z}) = \sum_{k=1}^K \pi_k \mathcal{N}(\bv{z} ; \bm{\mu}_k, \bm{\Sigma}_k) \ ,
\end{equation}
where $\bg{\theta}$ denotes the set of parameters $\bg{\theta} = \{ \pi_1,\ldots,\pi_K \in [0,1], \bm{\mu}_1,\ldots \bm{\mu}_K \in \mathbb{R}^d, \bm{\Sigma}_1,\ldots,\bm{\Sigma}_K \in \mathbb{S}_{+} ^d \}$ with $\mathbb{S}_{+} ^d$ denoting the set of symmetric, positive definite matrices of size $d \times d$. 
Here, the collection of mean vectors and covariance matrices are referred to succinctly as $\Ub = (\mub_1,\ldots,\mub_K)$, $\Sgmm = (\Sigmab_1,\ldots,\Sigmab_K)$, respectively. 
The procedure begins by determining the mean vectors $\Ub$ as local minima of $- \log \phi(\bv{z})$, the negative log of the unnormalized posterior PDF $\phi(\zb) = p(\data \giv \bv{z}) p(\zb)$ through global optimization. 
To discover multiple local minima  well-known global optimization methods such as simulated annealing and genetic algorithms could be used.
In the proposed method, repeated local optimization is employed in the form of multistart gradient descent with initial locations given by low-discrepancy Sobol samples of the domain. This approach, while not the most efficient, was chosen for robustness as evaluation of the log likelihood and its gradient is computationally inexpensive for the Bayesian inverse problems considered in section \ref{sec:results}.


The global optimization stage results in a set of local minima $\bv{z}_1^*,\ldots,\bv{z}_K^*$ taken as the centers $\bg{\mu}_1,\ldots,\bg{\mu}_K$ of a Gaussian mixture model with $K$ components. To estimate the covariance matrix of each component, the Laplace approximation is employed, resulting in
\begin{equation}
	\bv{\Sigma}_i \approx \bv{H}_f^{-1} (\bg{\mu}_i) 
\end{equation}
where $f(\bv{z}) = - \log \phi(\bv{z})$ and $\bv{H}_f(\bv{z})$ denotes the Hessian of $f$ evaluated at $\bv{z}$.  The LA uses a quadratic approximation of the log posterior to provide a Gaussian approximation at a mode or, equivalently, a local maximum-a-posteriori (MAP) estimate. It can also be viewed as an exact posterior arising from a local linearization of the model in a Bayesian inverse problem about the relevant mean vector \cite{immerImprovingPredictionsBayesian2021}. 
Note that the LA reflects the local geometry of a mode and will not reflect non-Gaussian trends away from the local MAP estimate. 
In contrast, the VI approximation of the posterior proceeds by minimizing KL-divergence between the surrogate and true posteriors and considers the non-Gaussian trends around the local MAP estimates.

The reader may note that several difficulties may arise in the computation of the Hessian during the LA stage. The first is the poor scalability of forming and inverting the Hessian matrix, computations which require $\mathcal{O}(d^2)$ and $\mathcal{O}(d^3)$ operations, respectively, with $d$ being the dimension of the parameter space.
The proposed algorithm was designed with large scale machine learning problems in mind where both of these are infeasible to carry out. 
The second issue is poor conditioning of the Hessian which could potentially have some zero eigenvalues to within machine precision. Both the scalability and conditioning issues can be addressed by considering various approximations to the Hessian matrix, often used in 2nd-order optimization methods. 
Correlation information can be limited by, for example, considering diagonal approximations \cite{daxbergerLaplaceReduxEffortless2021} of the Hessian as is also implicitly done in the Adam optimizer  \cite{kingmaAdamMethodStochastic2014}.  Diagonal approximations can be too restrictive for some problems, in which case alternative approximations can be used. For example Kronecker-factored approximations of the Fisher information matrix, (K-FAC) \cite{ritterScalableLaplaceApproximation2018,martensOptimizingNeuralNetworks2020} provide a block-diagonal approximation of the Hessian. Low-rank Hessian approximations offer another approach \cite{maddoxRethinkingParameterCounting2020} and can be combined with K-FAC \cite{leeEstimatingModelUncertainty2020}. Both diagonal and K-FAC approximations are guaranteed to be positive-semidefinite and positive definiteness can be ensured using the regularizing effect of a prior distribution \cite{daxbergerLaplaceReduxEffortless2021}.

The global optimization and LA stages provide us with Gaussian approximations at a number of modes of the posterior. A important question remains of how many modes to approximate as components of the GMM or, equivalently, when to stop the global search for modes of the posterior. 
We rely on the global search to find all the relevant modes and then the task is to select which modes are needed to represent the posterior accurately.
Model selection methods like automatic relevance determination \cite{bishop2006pattern}  and Akaike information criterion (AIC)/Bayesian information criterion (BIC) \cite{gelman1995bayesian}  could be used to determine which modes contribute significantly.
We found that the main issue is that some modes may be found multiple times and/or are not distinct.
Thus there is a need for reducing the collection of discovered local minima $\bm{\mu}_1,\ldots,\bm{\mu}_K'$ to a set of distinct means $\bm{\mu}_1,\ldots,\bm{\mu}_K$ comprising the centers of the GMM components. It is assumed that modes represent distinct possible values for the unknown parameters and are well-separated.  A greedy algorithm is used to determine a distinct subset of modes among those found by global optimization and proceeds by iterating through $\mub_1,\ldots,\mub_K'$. 
For each component $k$, the null hypothesis $H_0$ that $\bv{z}^*$ belongs to component $k$, i.e., $\bv{z}^* \sim \mathcal{N}(\bv{z}\giv \bm{\mu}_k, \bm{\Sigma}_k)$ is considered for the purpose of carrying out a significance test. 
Letting $D_M (\bv{z}^*,\mathcal{N}(\bv{x}\giv \bm{\mu}_k, \bm{\Sigma}_k))$ be the Mahalanobis distance between $\bv{z}^*$ and the local Gaussian distribution, the significance test is expressed as
\begin{equation}
	p_k = p(D_M (\bv{z}^*,\mathcal{N}(\bv{z}\giv \bm{\mu}_k, \bm{\Sigma}_k)) \geq d \giv H_0) = 1 - \chi^2(d,n)
\end{equation}
where $\bv{z}^*$ is taken as a new component if $p_k \geq t$ for each $k=1,\ldots,K'$ where $t$ is some chosen threshold. This takes the covariance matrix of previously determined modes into account.

The last step involves determination of the unknown weights, $\pi_k$ for $k=1,\ldots,K$.  This is carried out by first considering the constrained minimization of the weighted $L_2$ norm
\begin{equation}
	\argmin_{\bar{\pib}} \int w(\zb) \left\{ \phi(\zb)  - \sum_{k=1}^K \tilde{\pi}_k \mathcal{N}(\zb \giv \mub_k,\Sigmab_k)  \right\}^2 \d{\zb} \hspace{3mm} \text{s.t.} \  \tilde{\pi}_k \geq 0
\end{equation}
where $\tilde{\pib} = \left( \tilde{\pi}_1,\ldots,\tilde{\pi}_K \right) $ are the unnnormalized component weights, i.e. $\pi_k = c \tilde{\pi}_k$ for $k=1,\ldots,K$, and $w(\zb) = \sum_{k=1}^K \omega_k \mathcal{N}(\bv{z} \giv \bm{\mu}_k, \bm{\Sigma}_k)$ with $\omega_k = 1/K$ for $k=1,\ldots,K$ is the weighting function. The problem is formulated in terms of unnormalized weights since $\phi(\zb)$ is the unnormalized posterior with $\int \phi(\zb) \d{\zb} \neq 1$. The weighting function $w$ is chosen to limit the $L_2$ discrepancy to the region of support of the GMM which lies near the modes and is taken to be a GMM whose components have mean vectors and covariance matrices that match those of the GMM approximation. Since the weights of the GMM are unknown, equal weights $\omega_k = 1/K$ are used for $w$ which can be thought of as choosing the maximum entropy categorical distribution over the components. The weighted $L_2$ norm is approximated via Monte Carlo by sampling $N$ points from $w(\zb)$ and forming a sum of squared residuals resulting in the constrained least squares problem
\begin{equation}
		\argmin_{\bg{\bar{\pi}}} \sum_{i=1}^N \left\{	\phi(\bv{z}_i) - \sum_{k=1}^K \tilde{\pi}_k \mathcal{N}(\bv{z}_i \giv \bm{\mu}_k, \bm{\Sigma}_k)   \right\}^2 \hspace{3mm} \text{s.t.} \  \tilde{\pi}_k \geq 0 
\end{equation}
for the unnormalized weights $\tilde{\pi}_1,\ldots,\tilde{\pi}_K$. Letting $Z = \sum_{k=1}^K \tilde{\pi}_k$, we can form the normalized approximation to $p(\bv{z})$ as $q_{\bg{\theta}}(\bv{z}) = \sum_{k=1}^K \pi_k \mathcal{N}(\bv{z} \giv \bm{\mu}_k, \bm{\Sigma}_k)$ where $\pi_k = \tilde{\pi}_k/Z$. 
It is noteworthy that GOLA also gives an estimate, $Z$, of the Bayesian model evidence \cite{sivia2006data}, a crucial quantity in Bayesian model selection and model averaging \cite{wasserman2000bayesian}, without further likelihood/posterior evaluations which involve potentially costly forward model simulations. \cite{beck2010bayesian}The Global Optimization with Laplace Approximations (GOLA) method can be  summarized in 4 steps:
\begin{center}
	\fbox{
		\begin{minipage}{0.8\textwidth}
			\textbf{GOLA algorithm}
			\begin{enumerate}
				\item Perform global optimization to obtain local minima taken as potential centers $\mub_1,\ldots,\mub_{K'}$ of a GMM.
				\item Apply greedy algorithm based on a Mahalanobis $p$-test to obtain distinct modes $\mub_1,\ldots,\mub_{K}$ where $K \leq K'$.
				\item Use Hessian  of $- \log \phi$ at each mode, calculated in step $1$, to form a Laplace approximation $\mathcal{N}(\mub_i ,\bv{H}_{-\log \phi}^{-1} (\bg{\mu}_i) )$.
				\item Carry out the constrained quadratic optimization problem $\argmin_{\bg{\pi}} \sum_{i=1}^N \left\{	\phi(\bv{z}_i) - \sum_{k=1}^K \tilde{\pi}_k \mathcal{N}(\bv{z}_i \giv \bm{\mu}_k, \bm{\Sigma}_k)   \right\}$ such that $ \tilde{\pi}_k \geq 0$ for $k=1,\ldots,K$ to obtain the weights.
				\item Normalize the weights $\pi_k = \tilde{\pi}_k/Z$, $k=1,\ldots,K$ where $Z = \sum_{k=1}^K \tilde{\pi}_k$ to obtain a GMM approximation .
			\end{enumerate}
		\end{minipage}
	}
\end{center}

Note that performing VI with a GMM surrogate results in an optimization problem on a parameter space of dimension $\mathcal{O}(d^2)$, where $d$ is the dimension of the model parameter space. This becomes infeasible for large-scale ML models. Furthermore, the loss function is typically nonconvex, several optimizations are likely needed to avoid poor local minima. On the other hand, initializing VI with the global optimization procedure typically starts closer to the optimal solution so that fewer VI iterations are required for convergence.
GOLA also carries out the local optimization problems in $\mathbb{R}^d$ instead of $\mathbb{R}^{\mathcal{O}(d^2)}$ replacing several high-dimensional optimization problems with a multitude of $\mathcal{O}(K)$ of lower-dimensional problems. For a given number of components $K$, repeated, and potentially parallel, optimizations in $\mathbb{R}^d$ are seen to be more efficient in practice.

\section{Numerical Investigations}\label{sec:results}
We will first carry out robustness and scalability studies to investigate the general performance of the GOLA method. 
Robustness is gauged using an approach based on applying global sensitivity analysis to an ensemble of posterior PDFs with different characteristics. 
Scalability is measured by looking at procedure timings across an ensemble of tests. The final experiments subsection presents an application of GOLA to a physics-based exemplar in structural dynamics.

\subsection{Robustness}\label{sec:robustness}
A standard approach for validating an approximation procedure is to apply it to a canonical test problem where the true solution is known such that a measure of approximation error can be accurately obtained. Evaluating the performance of the proposed method on one or a small number of test applications may not provide an understanding of the procedure's weaknesses or how its robustness depends on particular features of the application problem. Here, in an approach similar to Ref.\cite{puyComprehensiveComparisonTotalorder2021}, variance-based sensitivity analysis is used to study the behavior of the method over an ensemble of synthetic test problems. Sensitivity analysis of the approximation error over this ensemble then provides a global summary of robustness and the factors that it depends on. 

Each test consists of applying the GOLA procedure to a randomly generated GMM defined by a set of parameters which impact the complexity of the posterior PDF. These parameters will define the input factors $X_1,\ldots,X_k$ for the sensitivity analysis while the model output $Y = f(X_1,\ldots,X_k)$ is taken to be the accuracy of the resulting GOLA approximation of the GMM generated accordingly.  The first order and total order Sobol sensitivity indices $S_i$ and $S_{T_i}$ \cite{saltelliVarianceBasedSensitivity2010}, respectively, are used here to measure sensitivity of $Y$. Intuitively, the first order index $S_i$ measures how much of the total variance $\var(Y)$ of $Y$ is due to the effect of factor $X_i$ alone. The total order index $S_{T_i}$ measures how much of $\var(Y)$ is due to first order and higher order interactions of $X_i$ with all other factors, i.e., how $X_i$ interacts with each possible combination of other factors. Mathematically, these two indices are defined by
\begin{align}
	S_i  &= \frac{\var_{X_i}(\ex_{\bv{X}_{\sim i}}(Y \giv X_i))}{\var(Y)} \label{eq:first-order} \\
	S_{T_i} &= 1 - \frac{\var_{\bv{X}_{\sim i}}( \ex_{X_i}(Y \giv \bv{X}_{\sim i}))}{\var(Y)} \label{eq:total-order}
\end{align}
where $\Xb_{\sim i}$ denotes all factors \textit{but} $X_i$. Note that in the definition of the first order index Eq. \ref{eq:first-order}, the inner expectation is over all other factors $\Xb_{\sim i}$  with $X_i$ fixed and the outer variance accounts for each possible value of $X_i$. The total order index is defined similarly.

\begin{figure}[h]
	\centering
	\includegraphics[width=0.9\textwidth]{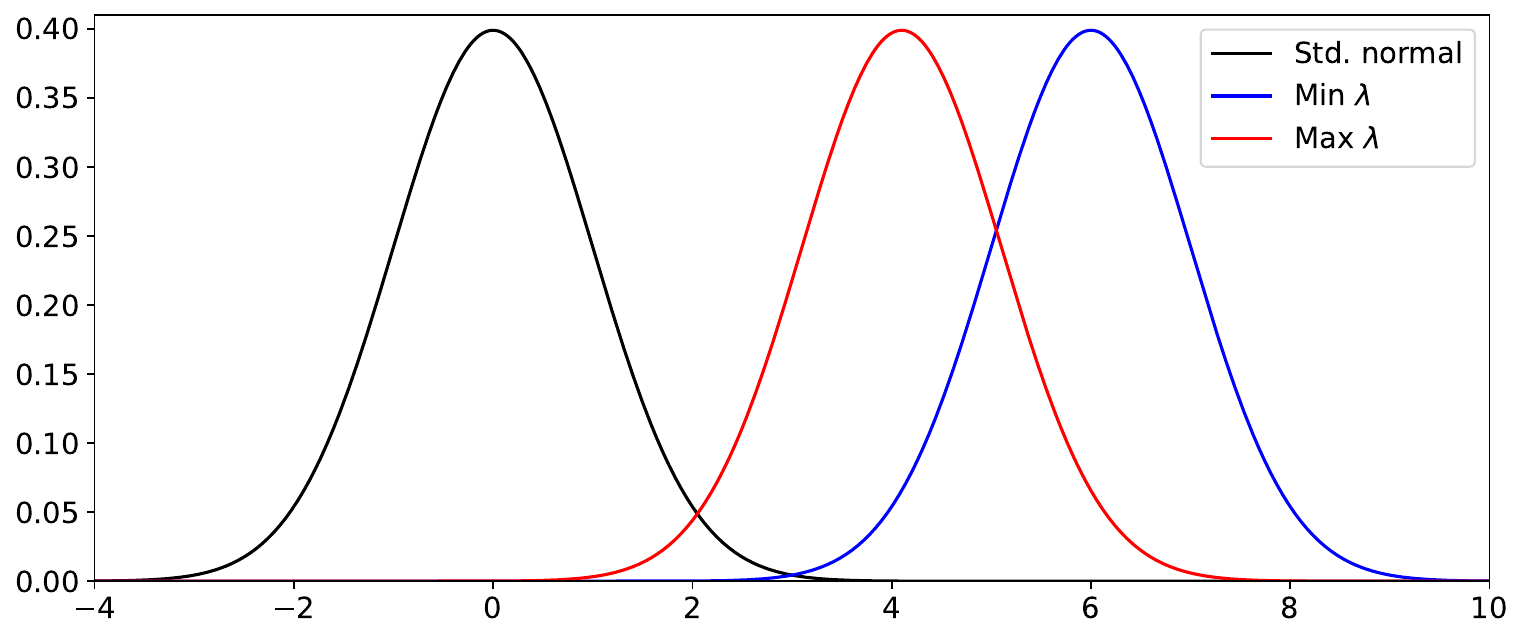}
	\caption{Visualization of lower and upper bounds for maximal overlap $\lambda$ listed in Table \ref{table:factor-distributions-1}. The overlap between the black and blue distributions is $10^{-4}$ while the overlap between black and red is $10^{-2}$.
}
	\label{fig:standard-normal-overlap}
\end{figure}

The factors defining the GMM test problems are described below:
\begin{itemize}
	\item \textit{Dimension:} As the dimension $d$ of model parameter space grows, the volume of the domain for the global optimization stage becomes larger yielding smaller probabilities for starting local searches in the basins of attraction of local minima. 
	\item \textit{Number of mixture components:} The number of mixture components $M$ reflects the number of modes in the true posterior that should be obtained through global optimization.
	\item \textit{Distribution of mixture weights:} A multiplicative decay factor $\omega$ is introduced which controls the variance in magnitude across mixture weights by $\pi_{k+1} = \frac{1}{\omega} \pi_k$ for $k=1,\ldots, K$. A large decay factor leads to a more uneven distribution of weights and the existence of less significant modes which are more difficult to locate.
	\item \textit{Correlation coefficient:} The correlation coefficient $c$ defines the off-diagonal entries of a mixture component's covariance matrix and affects the shape of a mode. As the correlation coefficient increases, probability mass is distributed in a non-isotropic manner leading to smaller basins of attraction and slower convergence of the gradient descent algorithm. Herein, we use a constant correlation coefficient $c$ across all components and parameter pairs.
	\item \textit{Overlap between components:} Overlap between GMM components increases non-Gaussian trends in the local posterior modes decreasing the effectiveness of the Laplace approximation and potentially making it harder to locate distinct local minima. While measuring the overlap between Gaussian mixture model components can be challenging \cite{{MeasuringComponentOverlapping},{nowakowskaTractableMeasureComponent2014}}, here the overlap $\lambda$ between distributions $p_1,p_2$ is taken to be the Dice metric
	\begin{equation}\label{eqn:overlap-measure}
		\lambda(p_1,p_2) = \frac{2 \int p_1(\bv{x}) p_2(\bv{x}) \d{\bv{x}}}{\int p_1^2(\bv{x}) \d{\bv{x}} +  \int p_2^2(\bv{x}) \d{\bv{x}}}
	\end{equation}
	which has a closed-form expression for Gaussian distributions  \cite{luMultivariateMeasuresSimilarity1989}. It is difficult to construct a mixture distribution such that all pairwise overlaps between the components have the same value. Hence, we take $\lambda$ to represent the maximal possible overlap between components such that at least one pair have overlap $\lambda$.
\end{itemize}
The five input factors defined above as well as the distributions over which they can vary are listed in Table \ref{table:factor-distributions-1}.
\begin{table}[h!]
	\centering
	\begin{tabular}{lll}
		\toprule
		Parameter     & Description     &  Distribution \\
		\midrule
		$d$ & Dimension  &  $\mathcal{U}\{ 2, 10 \}$  \\
		$M$     & Number of mixture components &  $\mathcal{U}\{ 2, 4 \}$     \\
		$\omega$   & Exponential decay factor across weights    & $\mathcal{U}[1, 2]$  \\
		$c$   & Correlation coefficient    & $\mathcal{U}[0, 0.7]$  \\
		$\lambda$   & Maximum overlap between components   & $\mathcal{U}[10^{-4},10^{-2}]$ \\
		\bottomrule
	\end{tabular}
	\caption{Robustness analysis factors and their distributions}
	\label{table:factor-distributions-1}
\end{table}
where $\mathcal{U}\{ a, a+k \}$ describes a discrete uniform distribution across values $a,a+1,\ldots,a+k$ while $\mathcal{U}[a, b]$ denotes a continuous one. The overlap measure $\lambda$ is difficult to visualize from formula \ref{eqn:overlap-measure} so the selected lower and upper bounds for the maximal overlap $\lambda \in [10^{-4},10^{-2}]$ between two standard, one-dimensional normal distributions are depicted in Figure \ref{fig:standard-normal-overlap}.

The accuracy function $f(d, M,\omega, c, \lambda)$ is defined as
\begin{equation} \label{eqn:sobol-function}
	Y = f(d, M,\omega, c, \lambda) = \jsdiv{\mathcal{G}(\pib,\Ub,\Sgmm)}{\mathcal{G}(\hat{\pib},\hat{\bm{\Ub}},\hat{\Sgmm})}
\end{equation}
where $\mathcal{G}(\pib,\Ub,\Sgmm)$ and $\mathcal{G}(\hat{\pib},\hat{\bm{\Ub}},\hat{\Sgmm})$ are the true and approximate GMMs, respectively, and $\jsdiv{\cdot}{\cdot}$ is the Jensen-Shannon divergence (JSD)
\begin{equation}
	\jsdiv{p}{q} = \frac{1}{2} \kldiv{p}{m} + \frac{1}{2} \kldiv{q}{m}; \ \ m=\frac{1}{2}(p+q)
\end{equation}
bounded in $\left[0,  \log(2)\right]$.  The JSD is rescaled to $\left[0, 1\right]$ to provide a normalized measure of the difference between two distributions and is estimated using Monte Carlo integration.  To compute the sensitivities, estimators from Refs  \cite{{saltelliVarianceBasedSensitivity2010},{puyComprehensiveComparisonTotalorder2021}} were used as these demonstrate the best performance among a collection of estimators. These are described in more detail in \ref{appx:global-sensitivity}. Bootstrap confidence intervals were also computed according to Ref. \cite{archerSensitivityMeasuresANOVAlike1997}.

GOLA displayed robust performance by obtaining a near perfect fit in $98\%$ of cases in the ensemble of generated posterior PDFs. To get a sense of when the procedure starts to break down, the parameter distributions in  \ref{table:factor-distributions-1} were modified to increase the probability of obtaining more complex posterior PDFs. The modified distributions  resulting sensitivity indices are listed in Table \ref{table:factor-distributions-2} along with their bootstrap confidence intervals.
\begin{table}[h!]
	\centering
	\begin{tabular}{lllll}
		\toprule
		Parameter     & Description     &  Distribution & $S$ & $S_T$\\
		\midrule
		$d$ & Dimension  &  $\mathcal{U}\{ 8, 9, 10 \}$ & $0.17 \pm 10^{-3}$ & $\mathbf{0.65} \pm 10^{-2}$  \\
		$M$     & No. of components &  $\mathcal{U}\{ 3, 4 \}$ & $0.13 \pm 10^{-3}$ & $0.30 \pm 10^{-3}$   \\
		$\omega$   & Weight decay    & $\mathcal{U}[1.3, 2]$ & $0.17 \pm 10^{-2}$ & $\mathbf{0.37} \pm 10^{-2}$ \\
		$c$   & Corr. coefficient    & $\mathcal{U}[0.1, 0.7]$  & $0 \pm 10^{-9}$ & $\mathbf{0.65} \pm 10^{-2}$\\
		$\lambda$   & Component overlap   & $\mathcal{U}[10^{-4},10^{-2}]$ & $0 \pm 10^{-9}$ & $0.02\pm 10^{-4}$ \\
		\bottomrule
	\end{tabular}
	\caption{Sensitivity analysis factors with refined distributions}
	\label{table:factor-distributions-2}
\end{table}
The first order effects are relatively small suggesting that most of the variability is due to interactions between factors. The three most significant total order effects are highlighted in bold and are associated with $d$, $\omega$, and $c$. These factors act in combination to form a distribution containing modes with basins of attraction whose volumes are small with respect to the total search region of the global optimization procedure. Random sampling is less likely to find these local minima. 

In summary, the performance of the global optimization stage of GOLA has the most significant impact on robustness. Therefore, posterior distributions with characteristics such as having modes which are less significant or display highly non-isotropic structure, present the largest challenges for the method, as with related schemes.

\subsection{Scalability}\label{sec:scalability}

\begin{figure}[h!]
	\centering
 	\includegraphics[width=0.8\textwidth]{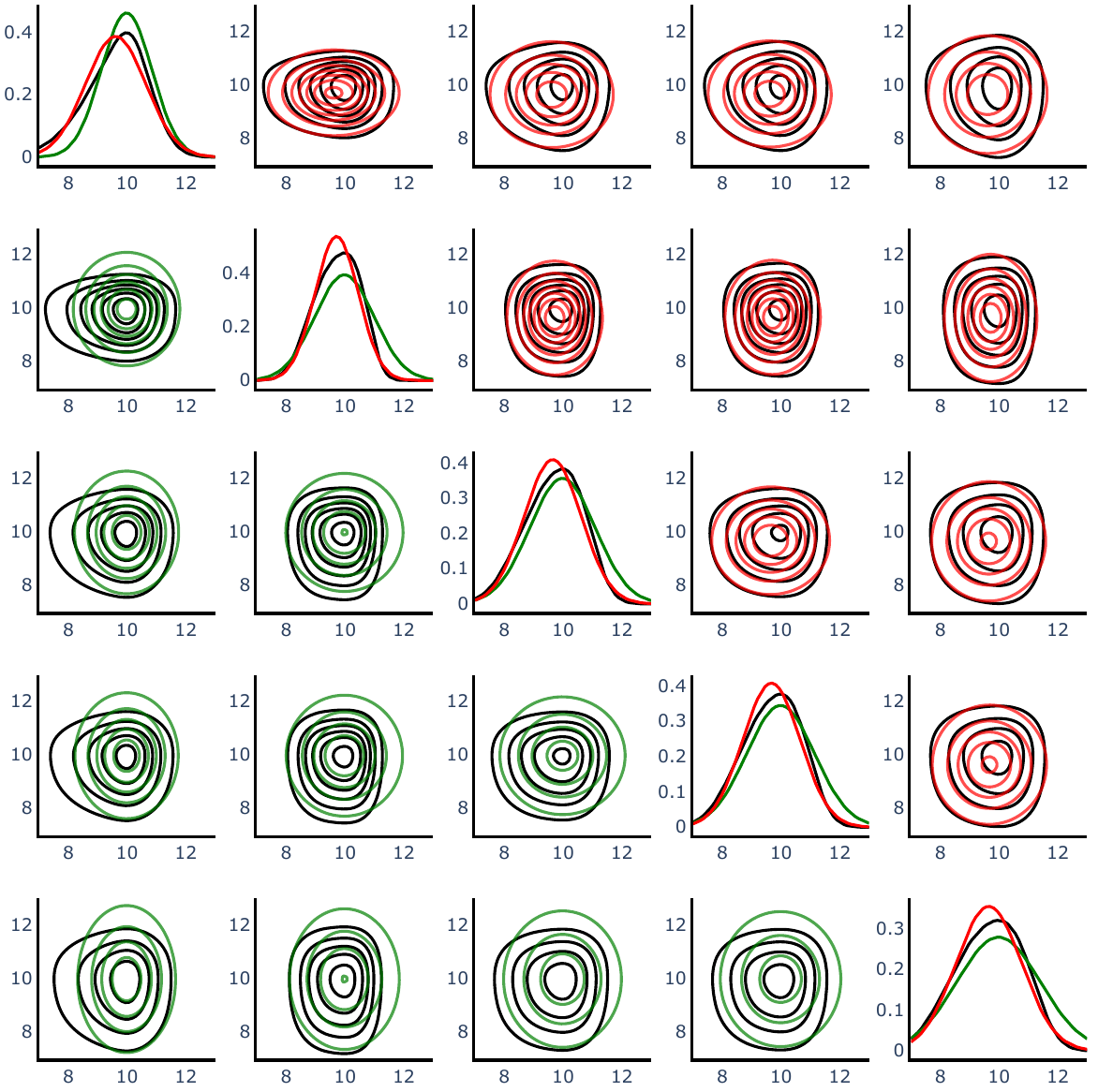}
	\caption{Two-dimensional marginal distributions for the five variables with the most skewness. Black represents the posterior while green and red represent the initial mixture model and VI-refined mixture model, respectively.
The panels on the diagonal compare all three for the self-correlations, while the lower panels compare the initial mixture to true and upper panels compare VI-refined to truth for the cross-correlations.}
	\label{fig:marginal-pyramid-plot}
\end{figure}

In this section, the  GOLA method is studied as an initialization procedure for VI. The analysis is carried out by constructing synthetic high-dimensional, multimodal posteriors through mixture distributions whose components display non-Gaussian trends. The scalability improvement gained through mixture model initialization is examined by comparing the average runtime of randomly initialized VI with the warm-start version carried out using the proposed approximation procedure. To obtain non-Gaussian behavior, the following nonlinear transformation of the standard normal distribution $Z$ is used
\begin{equation}
	Y = l + \sigma F(Z); \hspace{5ex} F(Z) =  \frac{\sinh((\arcsinh(Z)+s)t)}{2 \sinh(\arcsinh(Z)t)}
\end{equation}
which results in a random variable $Y$ with a Sinh-arcsinh distribution \cite{jonesSinharcsinhDistributions2009}. The parameters $l$ and $\sigma$ control the mean and variance while $s,t$ impact skewness and kurtosis. A non-Gaussian, multimodal distribution is constructed by forming a mixture model where each $n$-dimensional component is given by  a factorized product of $n$, $1$-dimensional Sinh-arcsinh distributions.
To gain intuition for how the GOLA and VI approximations differ, a $15$-dimensional synthetic posterior was generated randomly. GOLA was used to form an initial approximation of the distribution and subsequently refined by carrying out VI. The five posterior variables with the largest skewness values were selected and all 1D and 2D marginals from each choice of two parameters are plotted in Figure \ref{fig:marginal-pyramid-plot}. 

Looking at the 1D marginals, we can see that while VI further refines the approximation by modifying both the mean and covariance variational parameters, the initial mixture model is close to the final solution given by VI. The refinement is a consequence of the LA and KL-divergence representing different objective functions. The LA is based on local geometry while KL-divergence is a global measure of similarity.

\begin{figure}[h]
	\centering
	\includegraphics[width=0.6\textwidth]{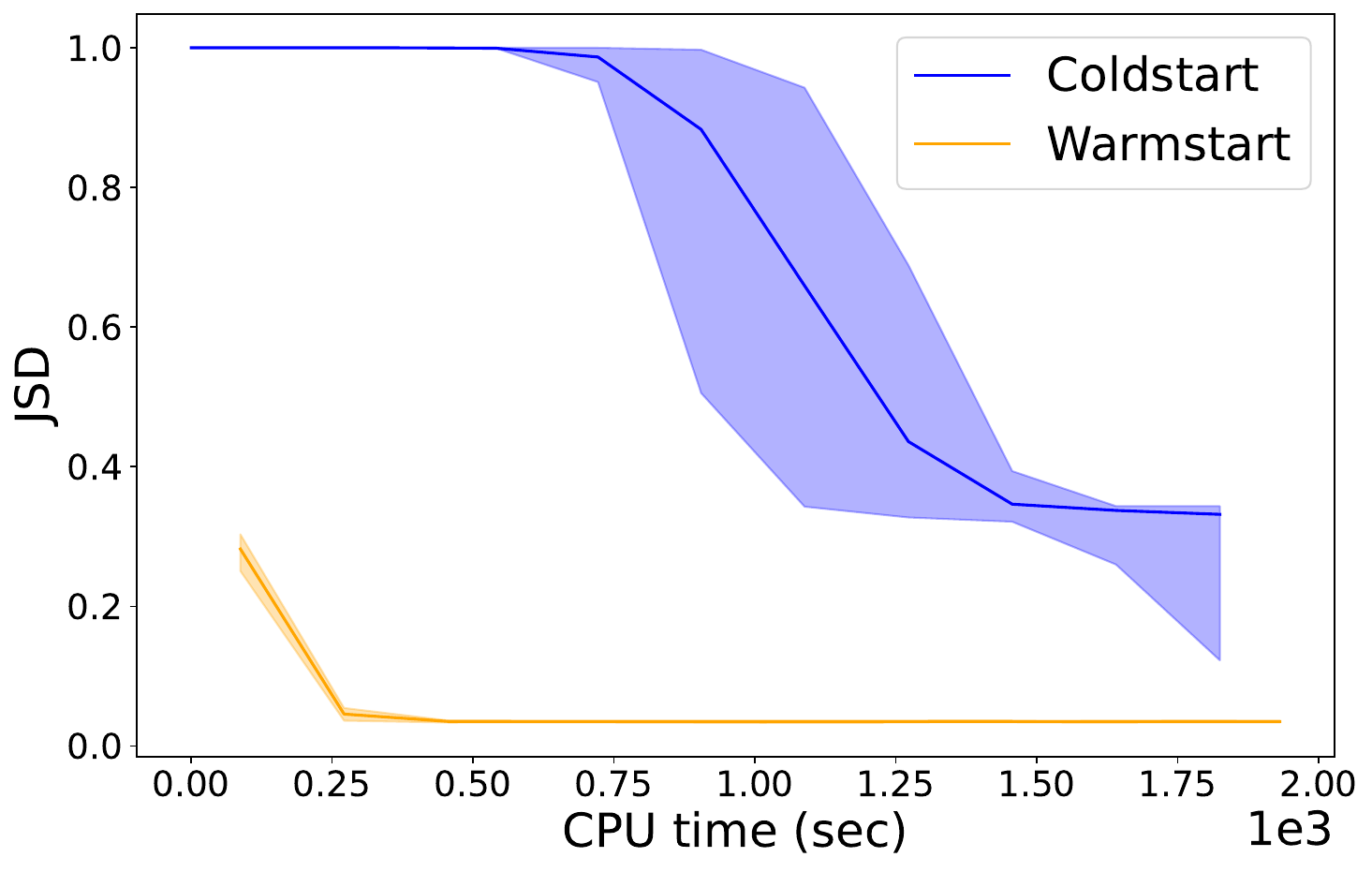}
	\caption{Mean (solid line) and $95\%$ confidence interval for cold-start (blue) and warm-start (orange) realizations for 15-dimensional problem}
	\label{fig:scalability-var}
\end{figure}

Next, scalability in terms of computation cost verus problem dimension is compared between randomly initializing VI (cold-start) versus initializing it with the GOLA procedure (warm-start). 
In both cases, the approximation accuracy is plotted as a function of elapsed CPU runtime with the JSD taken as the error metric. 
This scalability analysis was carried out on a machine with a 2.3 GHz Quad-Core Intel Core i7 processor and 32 GB of 3733 MHz memory. The synthetic posterior considered has two non-Gaussian modes but the dimension of the distribution is varied through 15, 30, and 60. Because of the stochastic behavior inherent in VI and the initialization procedures, multiple runs of both the cold-start and warm-start procedures were carried out.
\begin{figure}[h]
	\centering
	\includegraphics[width=1.0\textwidth]{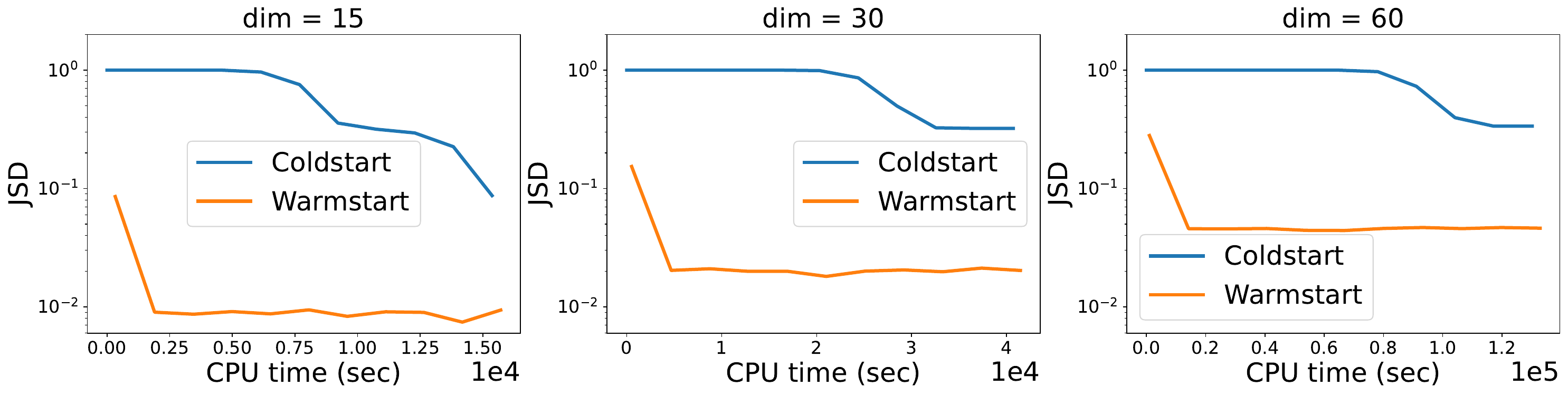}
	\caption{JSD versus elapsed CPU time of cold-start (blue: randomly initialized GMM) versus warm-start (red: GMM initialized with global optimization)
}
	\label{fig:scalability}
\end{figure}
The computational expense versus accuracy for each problem dimension is plotted in Figure \ref{fig:scalability}. To account for the multiple realizations carried out, the gradient descent in SVI was divided into epochs. Each point in the plot represents the sum of the total CPU time at a given epoch along with the minimum JSD value at that epoch taken across all realizations. Hence, the graph represents the best accuracy achieved with respect to the total computation time. In each case, the warm-start procedure accelerates convergence by a factor of at least 6. Furthermore, the warm-start procedure achieved a smaller overall JSD value. These trends are also visible in Figure \ref{fig:scalability-var} where the mean JSD and $95\%$ confidence intervals are plotted for 50 warm-start and cold-start runs for a 15-dimensional synthetic posterior.

In addition to the convergence benefits, it is also clear in Figures \ref{fig:marginal-pyramid-plot} and \ref{fig:scalability} that the initial GMM constructed by the GOLA procedure is a reasonable approximation to the true posterior.
This suggests that the LA may provide a cheaper alternative to VI in some cases.
Indeed in Reference \cite{daxbergerLaplaceReduxEffortless2021}, the authors show that LA is competitive with several standard approximate Bayesian inference procedures including Deep Ensembles \cite{lakshminarayananSimpleScalablePredictive2017} , mean-field Variational Bayes with Flipout \cite{blundellWeightUncertaintyNeural2015}, and cyclical stochastic-gradient Hamiltonian Monte Carlo \cite{zhangCyclicalStochasticGradient2020}. Additionally, the LA offers the smallest computational cost across all of these methods.

\subsection{A physics-based exemplar: multimodal structural dynamics}\label{sec:physics-application}

\begin{figure}[h]
	\centering
	\includegraphics[width=0.45\textwidth]{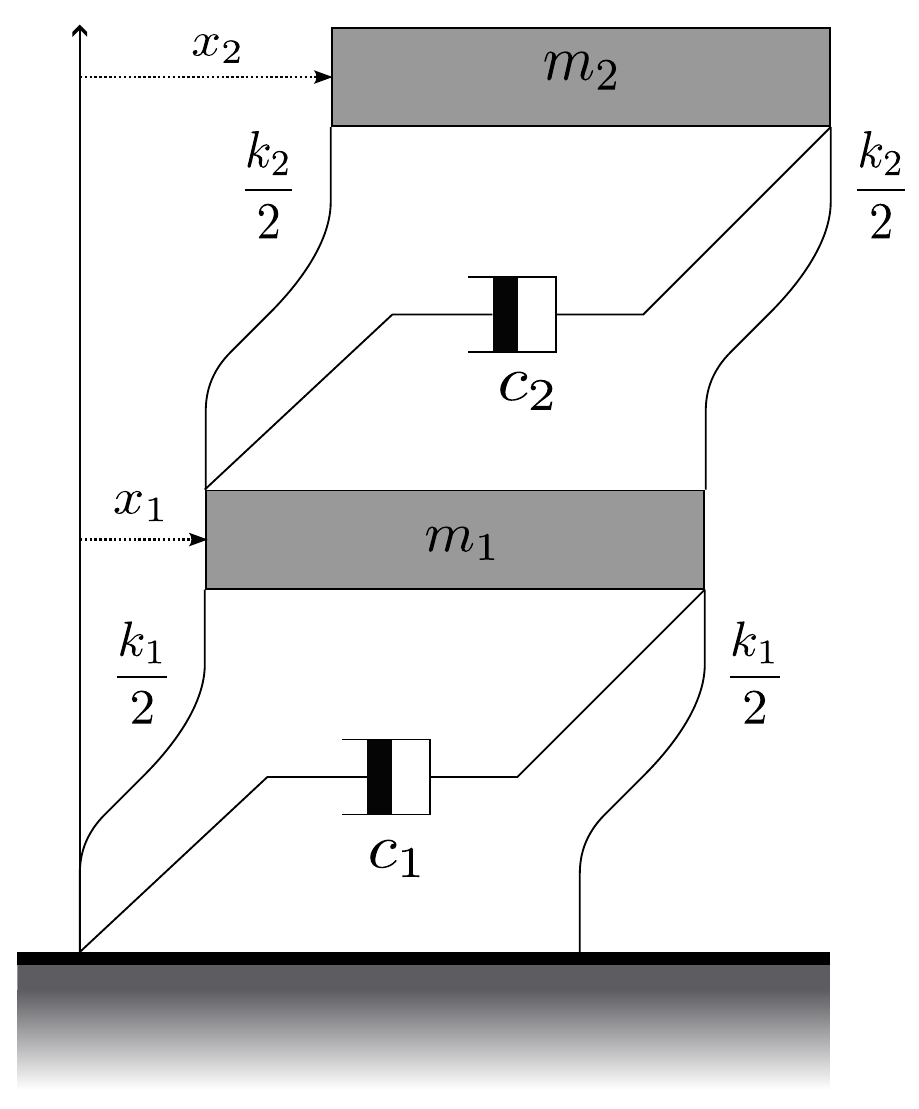}
	\caption{Two-story building modeled as a mass-spring-damper system \cite{augustiDynamicsStructuresTheory1996,adhikariDampingModelsStructural2000}.
Floor displacements are denoted by $x_i$, their masses are denoted as $m_i$.
The floor-wise stiffnesses are $k_i$ and the damping coefficients are $c_i$.}
	\label{fig:mass-spring-structure}
\end{figure}

To illustrate the practical value of GOLA for Bayesian model calibration, we consider a physical problem in which a two-story shear frame model is subjected to an initial excitation and formulate a Bayesian inverse problem for the unknown viscous damping.
The shear frame structure is depicted in Figure \ref{fig:mass-spring-structure}.
The mass is assumed to be concentrated at each floor and the beams are taken to be infinitely stiff with axial deformations neglected.
This leads to a highly idealized system whose physical degrees of freedom consist of the horizontal displacements $x_1$, $x_2$ of the floors from equilibrium. The constants $m_i$, $k_i$, $c_i$, $i=1,2$ define the mass, vertical beam stiffness and damping coefficients, respectively. Here, a two-dimensional Bayesian inverse problem for the unknown damping coefficients $c_1,c_2$ is considered for ease of visualization. The resulting likelihood is expensive to evaluate and exact gradient information is no longer available requiring the use of numerical derivatives in the GOLA procedure. Due to the cost of evaluating the likelihood, carrying out variational inference for this two-dimensional problem is computationally intensive and can benefit from application of GOLA.

The equations of motion can be written in matrix form as
\begin{equation}
	\bv{M} \ddot{\bv{x}} + \bv{C} \dot{\bv{x}} + \bv{K}\bv{x} = \bv{0}
\end{equation}
where $\bv{M}$ is a diagonal mass matrix, $\bv{C}$ is the matrix of viscous damping coefficients, and $\bv{K}$ is the stiffness matrix. The damping and stiffness matrices are given by
\begin{gather}
	\bv{C}
	=
	\begin{bmatrix}
		c_1 + c_2 & -c_2  \\
		-c_2 & 	c_2
	\end{bmatrix},
	\bv{K}
	=
	\begin{bmatrix}
		k_1 + k_2 & -k_2  \\
		-k_2 & 	k_2
	\end{bmatrix}
\end{gather}
This second-order system can be recast in state-space form as
\begin{gather}
        \dot{\ub} \equiv
	\frac{d}{dt}
	\begin{bmatrix}
		\bv{x} \\
		\bv{v}
	\end{bmatrix}
	=
	\begin{bmatrix}
		\bv{0} & \bv{I} \\
		-\bv{M}^{-1}\bv{K} & -\bv{M}^{-1}\bv{C} 
	\end{bmatrix}
	\begin{bmatrix}
		\bv{x} \\
		\bv{v}
	\end{bmatrix} 
        \equiv \Ab \ub
	\label{eqn:shear-frame-system}
\end{gather}
where $\bv{x},\bv{v}$ are the vectors of floor displacements and velocities, respectively, and $\ub = \left[ \xb  \hspace{2mm} \vb \right]^T$. 
The solution $\ub(t)$ at time $t$ with initial condition $\ub_0$ is given by the matrix exponential 
\begin{equation}
	\ub(t)  = e^{\Ab t}\ub_0 \label{eq:matrix-exp}
\end{equation}
where and $e^{\Ab t}$ is the matrix exponential of $\Ab t$ \cite{hartman2002ordinary}.

It is well known that inference of the damping coefficients is a difficult problem due to the complex relationship between damping forces and the other parameters of the system  \cite{augustiDynamicsStructuresTheory1996,adhikariDampingModelsStructural2000}.
Due to multiple resonances, Bayesian inference of the damping coefficients given sparse (in space and time) and noisy observations of the system results in a multimodal posterior.

\begin{figure}[h]
	\centering
	\includegraphics[width=0.8\textwidth]{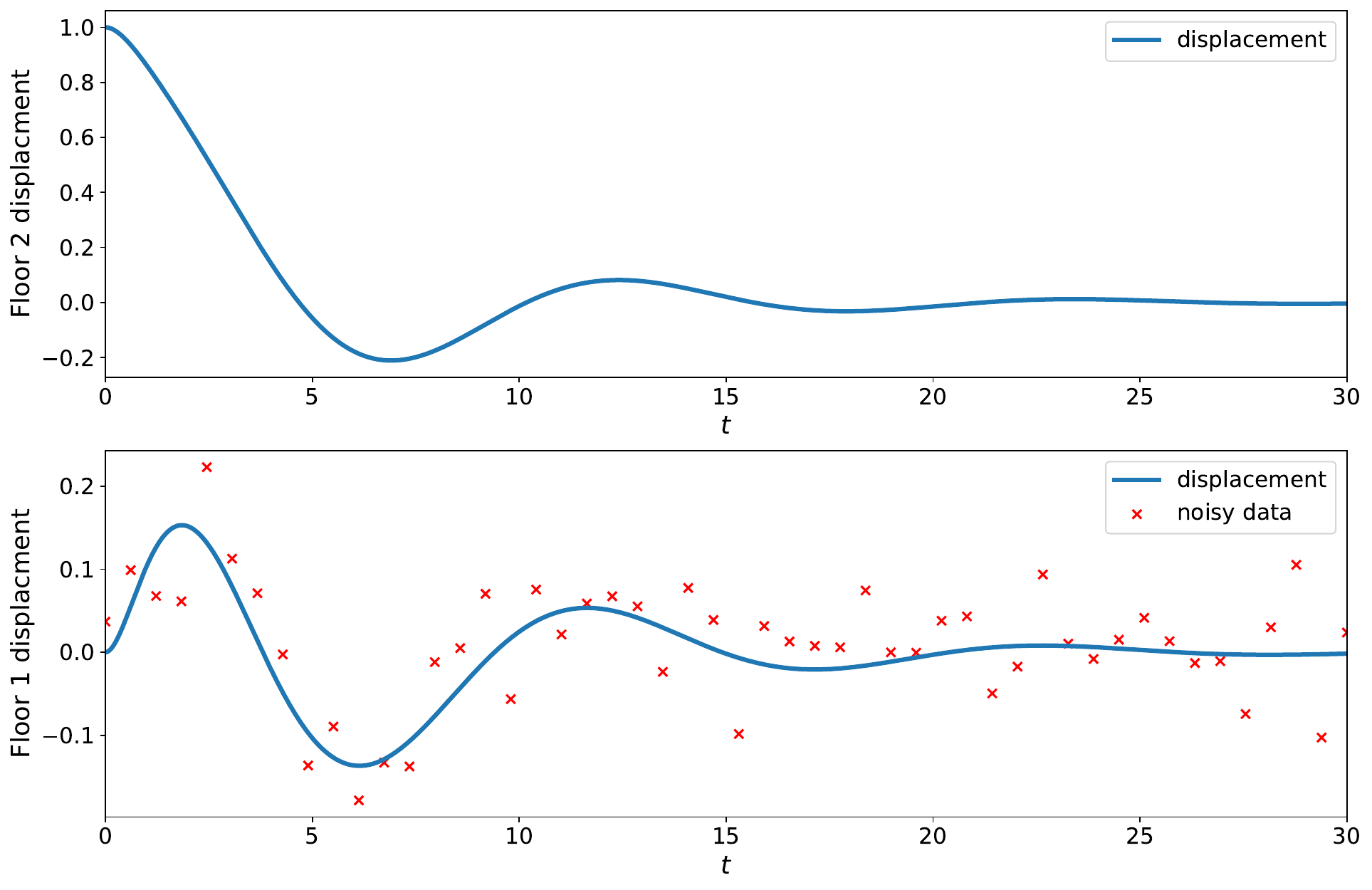}
	\caption{Displacement of both floors over a time interval of length 30 with noisy observations of the first floor (lower panel) after giving the second floor (upper panel) an initial displacement.}
	\label{fig:noisey-observations}
\end{figure}

\begin{figure}[h]
	\centering
	\includegraphics[width=0.8\textwidth]{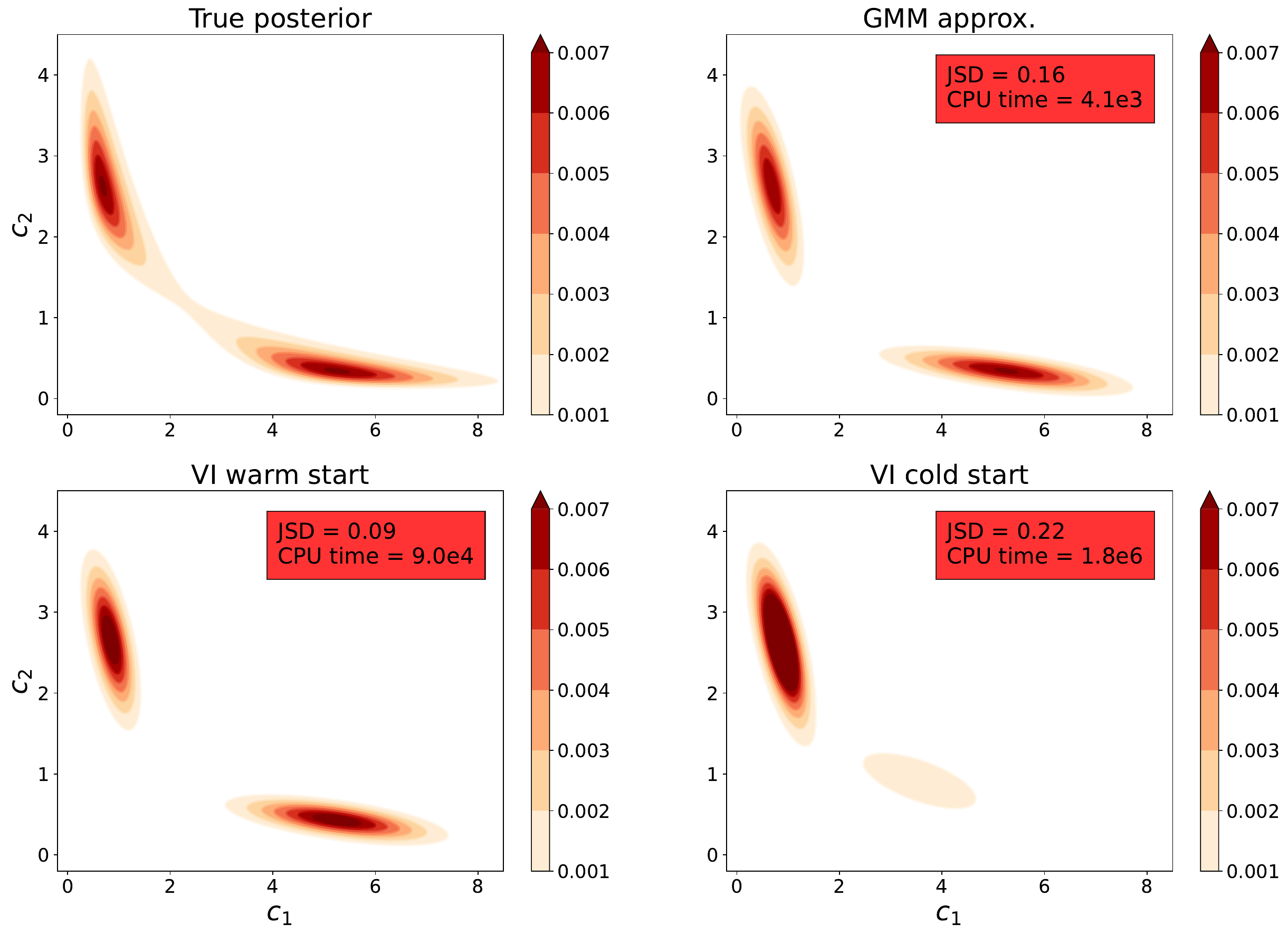}
	\caption{Contour plots of GOLA approximation, VI-refined (warm start) solution, and best case cold-start solution over 10 iterations. Also displayed are the minimum JSD values and corresponding wall-clock times}
	\label{fig:gola-vi-contour}
\end{figure}

In this example, the observations consist of noise-corrupted first floor displacements given an initial nonzero displacement of the second floor, with the measurement equation $y_i = \Hb \ub (t_i) + \epsilon$ used to model additive white (in time) Gaussian noise $\epsilon$ with assumed standard deviation $\sigma$. $\bv{H}$ is the linear observation operator that returns the first floor displacement from the state vector. The synthetic responses of the two floors, along with the available noisy observations, are displayed in Figure \ref{fig:noisey-observations}.
The $N_D$ observations of the first floor's displacement $\{ (t_i,y_i)\}_{i=1}^{N_D}$ under the assumption of independent Gaussian noise result in the following log likelihood function
\begin{equation}
	l(c_1,c_2) = \frac{1}{\sigma^2}\sum_{i=1}^{N_D} (y_i - \bv{H} \ub (t_i))^2 \ .
\end{equation}
with the prior taken to be uninformative. The state response at a given time instance of interest $t = t_i$ is given by equation \ref{eq:matrix-exp}.  For the inversion tasks, we assume that the initial condition vector, $\ub_0$, consisting of starting displacements and velocities, is known. In Figure \ref{fig:gola-vi-contour}, contour plots of the GOLA approximation, VI-refined solution (warm start), and randomly initialized VI solution are shown. The GMM approximation accurately captures the location and local geometry of the modes while missing a curved, low-probability region connecting the two modes of the posterior. To evaluate the quality of this approximation, the GOLA approximation is compared to the approximation obtained by refining this solution using VI, as well as carrying out VI with a random initial condition.  The minimum JSD values achieved along with corresponding wall-clock times are also displayed in Figure \ref{fig:gola-vi-contour}.

In this case, VI mostly refines the locations of the two components, resulting in better overlap with the posterior.
This is a result of minimizing the KL-divergence objective function which penalizes the existence of high-probability regions of the surrogate posterior in low-probability regions of the true posterior.
Randomly-initialized (cold start) VI frequently becomes stuck in local minima, providing inconsistent approximations as illustrated in Figure \ref{fig:gola-vi-contour}.
\begin{figure}[h]
	\centering
	\includegraphics[width=0.8\textwidth]{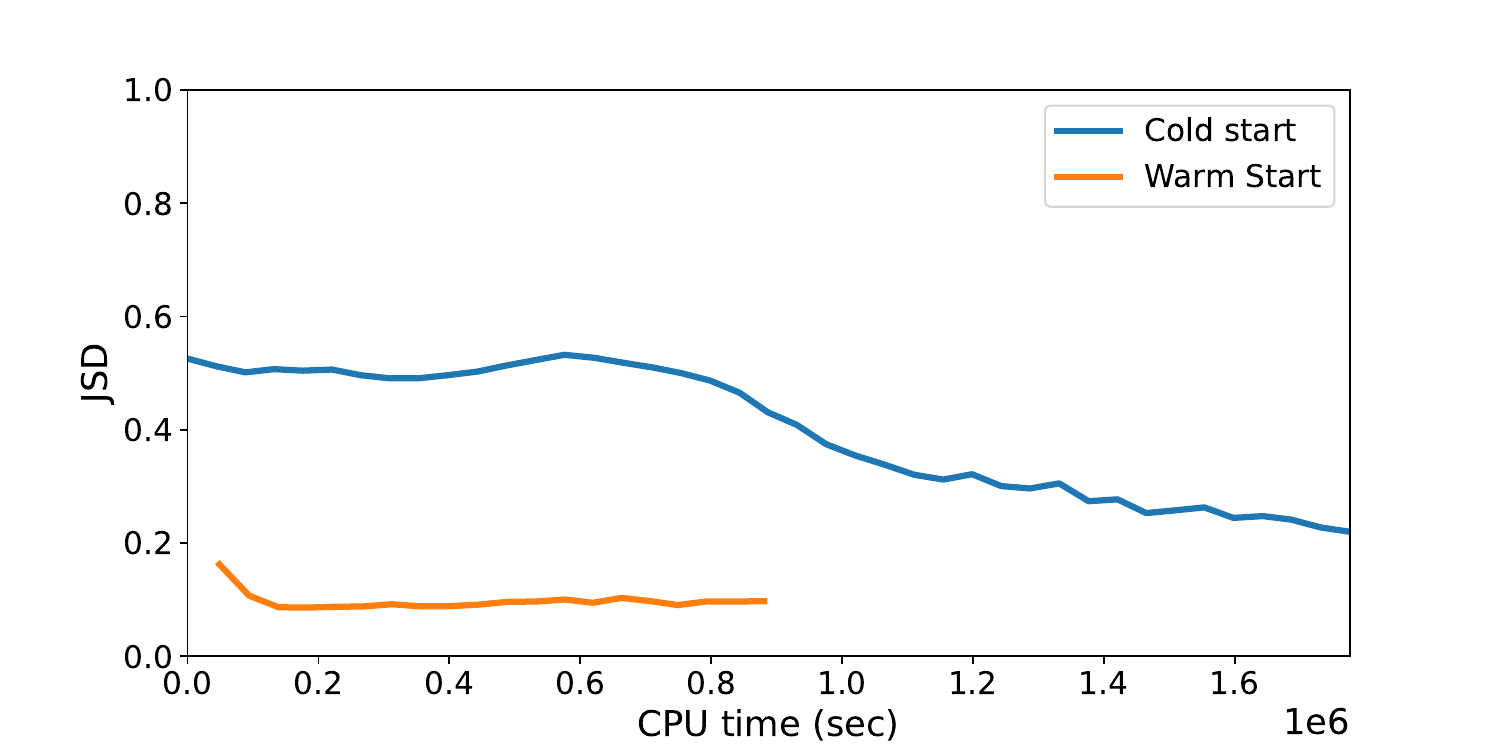}
	\caption{JSD versus elapsed wall-clock time for the GOLA initialized (warm start) and randomly initialized VI (cold start) where the warm start curve begins at the time elapsed after carrying out the GOLA procedure. The cold start cuve represents the best case over 10 independent runs.}
	\label{fig:osc-jsd-timing}
\end{figure}
The approximation accuracy as a function of elapsed wall-clock CPU time of the GOLA initialized and randomly initialized VI is given in Figure \ref{fig:osc-jsd-timing} where the GOLA initialized timings include the time required to obtain the GMM approximation.
The timings reflect a remarkably more rapid convergence to a high quality posterior approximation with the GOLA initialized VI.

\begin{figure}[h]
	\centering
	\includegraphics[width=0.9\textwidth]{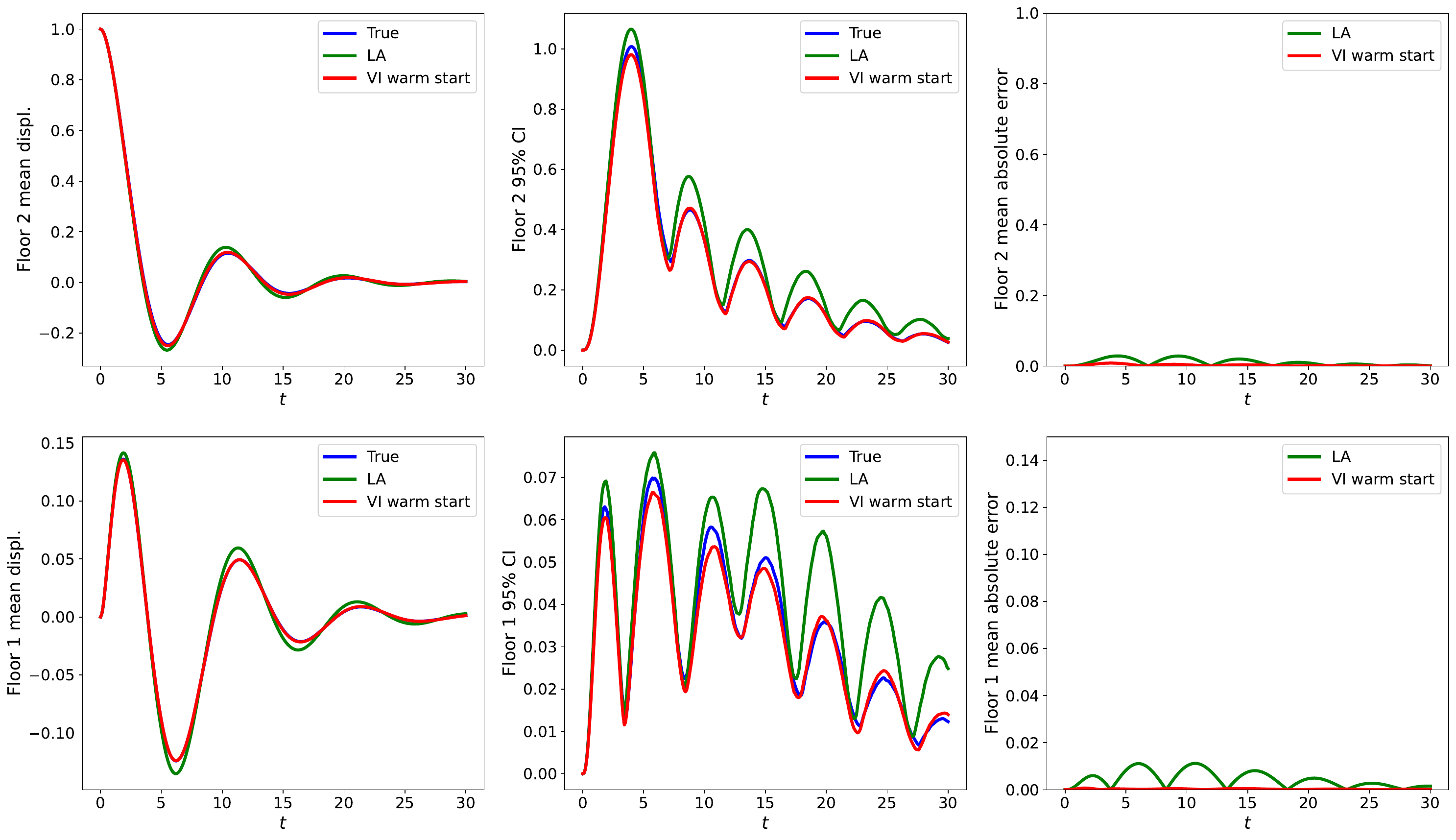}
	\caption{Mean displacement, $95 \%$ confidence interval, and absolute error with respect to true solution, as a function of time for first floor (bottom)and and second floor (top).}
	\label{fig:pushforward-posterior}
\end{figure}

\begin{figure}[h]
	\centering
	\includegraphics[width=0.7\textwidth]{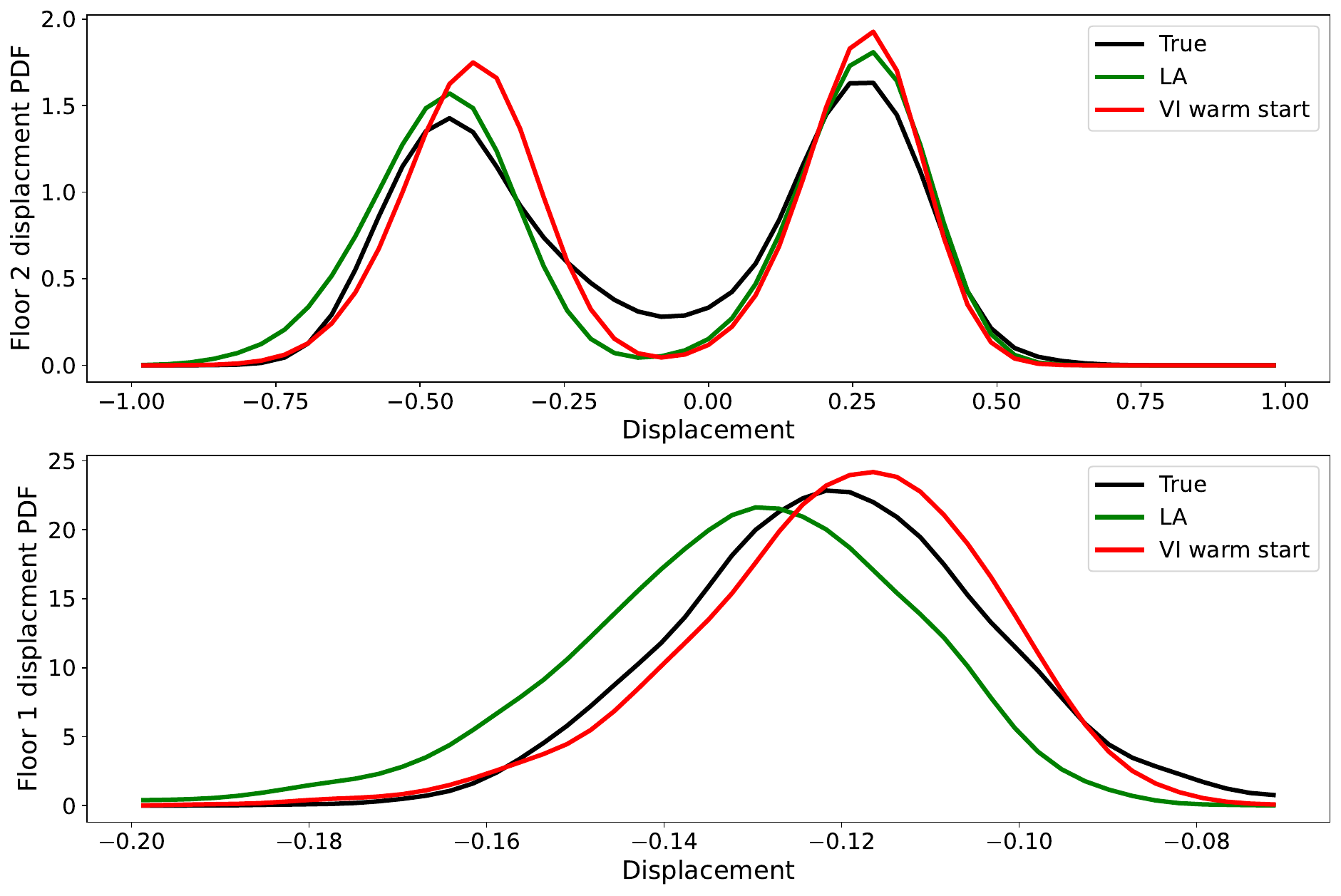}
	\caption{Pushforward posterior of the true posterior, GOLA approximation, and VI approximation at a particular time where $95\%$ confidence interval for the first floor's mean displacement is the widest.}
	\label{fig:pushforward-posterior-slice}
\end{figure}

The predictive distribution obtained by GOLA versus VI provides another significant point of comparison between the methods. Here, the distribution over model predictions is given by the pushforward posterior, obtained by propagating the parametric uncertainty through the model towards uncertain predicctions. Samples of the pushforward posterior are obtained by simulating the system on samples from the parameter posterior distribution.
The pushforward posteriors of the floor displacements using the true posterior along with both approximations are shown in Figure \ref{fig:pushforward-posterior}.
The mean trajectories of the GOLA and VI-refined pushforward posteriors both provide an accurate approximation of the true mean floor displacements with the Laplace approximations tending to slightly overestimate the uncertainty and the VI-refined distribution slightly underestimating the uncertainty. A similar behavior is seen when looking at Figure \ref{fig:pushforward-posterior-slice}, which shows the distribution of floor displacements at the particular time where the $95\%$ confidence interval for the first floor's mean displacement is the widest.

\section{Conclusions}

In this paper, we presented the GOLA algorithm, a scalable method for initializing VI on high-fidelity GMM surrogate posteriors. Multi-modal approximations are often needed as multi-modal posteriors are often encountered in inverse problems of nonlinear-in-parameter ML and physics-based models. We first showed that the procedure is robust over a wide range of possible true posterior distributions using a Sobol sensitivity analysis. It was seen that posterior features which affect the difficulty of finding local minima during global optimization are the biggest challenge for accuracy. Next, we established that the scalability of VI is greatly improved through our initialization procedure. In particular, the time required for VI to converge is significantly reduced in comparison to cold-start and this improvement increases with the underlying dimensionality of the true posterior distribution.

While investigating the scalability of the procedure, we also observed that the initial mixture model constructed by GOLA formed a reasonable approximation to the true posterior. This observation is corroborated by other work  that shows the LA performs well in Bayesian inference tasks when compared to other popular approaches such as VI and MCMC. Yet the best performing method for posterior estimation is typically problem dependent. This suggests that for some tasks, GOLA can provide a cheaper alternative to VI for GMMs that offers similar accuracy.

\section*{Acknowledgments}
This work supported by the Laboratory Directed Research and Development program  (Project 222361).
This paper describes objective technical results and analysis. Any subjective views or opinions that might be expressed in the paper do not necessarily represent the views of the U.S. Department of Energy or the United States Government.
Sandia National Laboratories is a multimission laboratory managed and operated by National Technology and Engineering Solutions of Sandia, LLC., a wholly owned subsidiary of Honeywell International, Inc., for the U.S. Department of Energy's National Nuclear Security Administration under contract DE-NA-0003525. This paper describes objective technical results and analysis. Any subjective views or opinions that might be expressed in the paper do not necessarily represent the views of the U.S. Department of Energy or the United States Government.

\clearpage

\bibliographystyle{unsrt}  

\clearpage
\appendix

\section{Global, variance-based sensitivity analysis} \label{appx:global-sensitivity}

Variance-based sensitivity analysis describes how the global variance of a function, $f(x_1,\ldots,x_k)$, usually thought of as a model response, can be attributed to combinations of the input factors $x_1,\ldots,x_k$.
The procedure is carried out by considering the inputs as random variables $X_1,\ldots,X_k$ and decomposing the total variance $\var(Y)$  of $Y=f(X_1,\ldots,X_k)$ as follows
\begin{equation} \label{eqn:function-decomposition}
	f = f_0 + \sum_i f_i(X_i) + \sum_i \sum_{j>i} f_{i j}(X_i, X_j) + \cdots + f_{1 2 \ldots k}(X_1,\ldots,X_k)
\end{equation}
where, under certain assumptions, the individual functions are defined by the following expectations 
\[
f_0 = \ex [Y], \ \ f_i = \ex_{\bv{X}_{\sim i}} [Y \giv X_i] - f_0, \ \ f_{i j} =  \ex_{\bv{X}_{\sim i j}} [Y \giv X_i] - f_i - f_j  - f_0, \ \ \ldots
\]
where the notation $\bv{X}_{\sim i}$ means all variables except $X_i$. Dividing a term $\var(f_{i_1,\ldots,i_s})$ by $\var(Y)$ yields the sensitivity index $S_{i_1,\ldots,i_s}$.  Taken together, the sensitivity indices satisfy the relation
\begin{equation}
	\sum_i S_i +  \sum_i \sum_{j>i} S_{i j} + \cdots + S_{1 2 \ldots k} = 1
\end{equation}
that shows the total variance is partitioned among the factors. Often, only the first and total order indices are computed which are defined by the formulas
\begin{align}
	S_i  &= \frac{\var_{X_i}(\ex_{\bv{X}_{\sim i}}(Y \giv X_i))}{\var(Y)} \\
	S_{T_i} &= 1 - \frac{\var_{\bv{X}_{\sim i}}( \ex_{X_i}(Y \giv \bv{X}_{\sim i}))}{\var(Y)}
\end{align}
The first order indices $S_i$ measures the variability due to factor $X_i$ alone while the total indices $S_{T_i}$ account for all possible interactions of $X_i$ with other factors. 

To estimate the first and total order sensitivity indices of $f(d, K,\omega, c, \lambda)$ with respect to each factor, sampling matrices $\bv{A},\bv{B} \in \mathbb{R}^{N\times k}$ are formed where $k$ is the number of factors and $N$ the number of samples of the random vector $(X_1,\ldots,X_k)$. An additional set of matrices $\bv{A}_{\bv{B}}^{(i)}$ is introduced for each $i=1,\ldots,k$ where all columns come from $\bv{A}$ except the $i^{\text{th}}$ column which is taken from $\bv{B}$.  The function $f$ is evaluated row-wise on these matrices to form vectors $f(\bv{A})$, $f(\bv{B})$, and $f(\bv{A}_{\bv{B}}^{(i)}), i=1,\ldots,N$ where all of these are in $\mathbb{R}^N$ \cite{saltelliGlobalSensitivityAnalysis2008} . Hence, the total number of model evaluations is $N(k+2)$. The following estimators were used to compute the sensitivity indices
\begin{align}
	S_i &= V(Y)^{-1}\frac{1}{N}\sum_{j=1}^N f(\bv{B})_j (f(\bv{A}_{\bv{B}}^{(i)})_j - f(\bv{A})_j) \\
	S_{T_i} &= V(Y)^{-1}\frac{1}{2N}\sum_{j=1}^N  (f(\bv{A})_j - f(\bv{A}_{\bv{B}}^{(i)})_j)^2
\end{align}
where $V(Y)$ is the total variance estimated as $V(Y)=\frac{1}{N} \sum_{i=1}^N(f(\bv{A})_i -f_0)^2$ with $f_0$ the sample mean of $f(\bv{A})$. These estimators have been shown to be the particularly efficient in terms of the variance of the provided estimates with respect to the number of samples \cite{{saltelliVarianceBasedSensitivity2010},{puyComprehensiveComparisonTotalorder2021}}. To provide uncertainty estimates on our first and total-order indices, bootstrap confidence intervals can be computed for sensitivity indices \cite{archerSensitivityMeasuresANOVAlike1997} by repeatedly resampling the initial set of sampling matrices $\bv{A}$,$\bv{B}$, and $\bv{A}_{\bv{B}}^{(i)}$ and computing the variance of estimates formed from resampled values.

\end{document}